\documentstyle[]{mn}
\newif\ifAMStwofonts
\def\ion#1#2{#1\,{\sevensize\uppercase\expandafter{\romannumeral#2}}} 

\def\micron{\hbox{$\mu$m}}
\def\mum{\hbox{\micron~}}
\def\lir{\hbox{$L_{\rm IR}$}}

\def\lsun{\hbox{L$_\odot$}}

\def\kms{\hbox{km s$^{-1}$}}
\def\brg{\hbox{Br$\gamma$}}
\def\trademark{{\ooalign{\hfil\raise.07ex\hbox{\tiny R}\hfil\crcr\mathhexbox20D}}}

\ifoldfss
  \ifCUPmtlplainloaded \else
    \NewTextAlphabet{textbfit} {cmbxti10} {}
    \NewTextAlphabet{textbfss} {cmssbx10} {}
    \NewMathAlphabet{mathbfit} {cmbxti10} {} % for math mode
    \NewMathAlphabet{mathbfss} {cmssbx10} {} %  "   "    "
  \fi
  \ifAMStwofonts
    \ifCUPmtlplainloaded \else
      \NewSymbolFont{upmath} {eurm10}
      \NewSymbolFont{AMSa} {msam10}
      \NewMathSymbol{\upi}     {0}{upmath}{19}
      \NewMathSymbol{\umu}     {0}{upmath}{16}
      \NewMathSymbol{\upartial}{0}{upmath}{40}
      \NewMathSymbol{\leqslant}{3}{AMSa}{36}
      \NewMathSymbol{\geqslant}{3}{AMSa}{3E}

      \let\leq=\leqslant 
      \let\geq=\geqslant 
    \fi
  \fi
\fi % End of OFSS

\ifnfssone
  \newmathalphabet{\mathit}
  \addtoversion{normal}{\mathit}{cmr}{m}{it}
  \addtoversion{bold}{\mathit}{cmr}{bx}{it}
  \newmathalphabet{\mathbfit} % math mode version of \textbfit{..}
  \addtoversion{normal}{\mathbfit}{cmr}{bx}{it}
  \addtoversion{bold}{\mathbfit}{cmr}{bx}{it}
  \newmathalphabet{\mathbfss} % math mode version of \textbfss{..}
  \addtoversion{normal}{\mathbfss}{cmss}{bx}{n}
  \addtoversion{bold}{\mathbfss}{cmss}{bx}{n}
  \ifAMStwofonts
    \ifCUPmtlplainloaded \else
      \UseAMStwoboldmath
      \makeatletter
      \new@mathgroup\upmath@group
      \define@mathgroup\mv@normal\upmath@group{eur}{m}{n}
      \define@mathgroup\mv@bold\upmath@group{eur}{b}{n}
      \edef\UPM{\hexnumber\upmath@group}
      \new@mathgroup\amsa@group
      \define@mathgroup\mv@normal\amsa@group{msa}{m}{n}
      \define@mathgroup\mv@bold\amsa@group{msa}{m}{n}
      \edef\AMSa{\hexnumber\amsa@group}
      \makeatother
      \mathchardef\upi="0\UPM19
      \mathchardef\umu="0\UPM16
      \mathchardef\upartial="0\UPM40
      \mathchardef\leqslant="3\AMSa36
      \mathchardef\geqslant="3\AMSa3E

      \let\leq=\leqslant 
      \let\geq=\geqslant 
    \fi
  \fi
\fi % End of NFSS release 1

\ifnfsstwo
  \DeclareMathAlphabet{\mathbfit}{OT1}{cmr}{bx}{it}
  \SetMathAlphabet\mathbfit{bold}{OT1}{cmr}{bx}{it}
  \DeclareMathAlphabet{\mathbfss}{OT1}{cmss}{bx}{n}
  \SetMathAlphabet\mathbfss{bold}{OT1}{cmss}{bx}{n}
  \ifAMStwofonts
    \ifCUPmtlplainloaded \else
      \DeclareSymbolFont{UPM}{U}{eur}{m}{n}
      \SetSymbolFont{UPM}{bold}{U}{eur}{b}{n}
      \DeclareSymbolFont{AMSa}{U}{msa}{m}{n}
      \DeclareMathSymbol{\upi}{0}{UPM}{"19}
      \DeclareMathSymbol{\umu}{0}{UPM}{"16}
      \DeclareMathSymbol{\upartial}{0}{UPM}{"40}
      \DeclareMathSymbol{\leqslant}{3}{AMSa}{"36}
      \DeclareMathSymbol{\geqslant}{3}{AMSa}{"3E}

      \let\leq=\leqslant 
      \let\geq=\geqslant 
    \fi
  \fi
\fi % End of NFSS release 2

\ifCUPmtlplainloaded \else
  \ifAMStwofonts \else % If no AMS fonts
    \def\upi{\pi}
    \def\umu{\mu}
    \def\upartial{\partial}
  \fi
\fi

\title[Infrared galaxies]{8--13 \mum spectroscopy of luminous and ultraluminous infrared galaxies}
\author[C. C. Dudley]
       {C. C. Dudley \\
        Institute for Astronomy, University of Hawaii, 2680 Woodlawn Dr., Honolulu, Hawaii 96822, USA }
\date{}
\pubyear{1999}

\begin{document}

\maketitle

\label{firstpage}

%\hfill\the\month /\the\day /\the\year

\begin{abstract}

New moderate-resolution mid-infrared spectroscopy at 10 \mum of 27
infrared galaxies is presented. The galaxies have been chosen
from three 60 \mum selected and one 12 \mum selected complete
flux-limited catalogs of galaxies; 17 of these sources have $
L_{\rm IR}($8--1000\micron$)
\geq 5\times 10^{11}$ \lsun.  A high-resolution spectrum
of the source Arp 299B1 is also presented.  Combining these
new results with previously published results, a nearly
complete 60 \mum selected flux limited subsample, with $
L_{\rm IR}($8--1000\micron$) \geq 1.6\times 10^{11}$
\lsun~ of 25 galaxies is defined. Within this subsample, it is found
that the dominant power source of infrared galaxies in the
luminosity range $1.6\times 10^{11} < L_{\rm
IR}($8--1000\micron$) < 10^{12} $
\lsun~ is massive star formation based on the clear
presence of the 11.3 \mum aromatic hydrocarbon emission
feature in the majority of the spectra.  Three of the five
ultraluminous infrared galaxies ($L_{\rm
IR}($8--1000\micron$) \geq 10^{12} $ \lsun) within this subsample show evidence that an
active galactic nucleus
provides an energetically important power source based on the detection of silicate absorption in their mid-infrared
spectra.  The physical basis of a possible
anti-correlation between the 11.3 \mum feature equivalent
width and infrared light to molecular gas mass ratios is
discussed.

\end{abstract}

\section{Introduction\label{ch5:int}}

This work is aimed at the question of what provides the
dominant power source of ultraluminous infrared galaxies, that
is galaxies with luminosities in the wavelength range
8--1000 \mum in excess of $10^{12}$ \lsun~($H_0$=75 km
s$^{-1}$ Mpc$^{-1}$).  This question is of interest because
it is possible that the power source of these galaxies is
qualitatively different from the dominant power source of most
infrared galaxies of lower luminosity.  The observational
approach of this work is the use of moderate resolution
8--13 \mum spectroscopy.  

In very broad strokes, the context of the debate over the
question addressed here can be summarized as 
an attempt, on the one hand, to explain ultraluminous infrared
galaxies as more powerful versions of starburst galaxies,
that is as super starbursts (Joseph \&
Wright~1985), where massive star
formation provides the power source, and on the other hand,
to demonstrate that ultraluminous infrared galaxies are
the precursors of quasars (Sanders et al. 1988) and are
powered by active galactic nuclei (AGNs).

The question has remained open to serious debate for
essentially one reason: dust.  Infrared galaxies by their
nature radiate the bulk of their energy through the 
degradation of the optical and UV photons emitted by the
primary power source into infrared photons emitted by dust
heated to a temperature of $\sim$50 K.  This has the
effect of erasing most of the commonly accepted
features used to distinguish the characteristics of
starbursts and AGNs.

Optical and, to a lesser degree, near-infrared studies are
particularly susceptible to this difficulty since spectroscopic signatures of AGNs in ultraluminous
infrared galaxies should only betray sources that are
relatively unobscured and therefore might not contribute
substantially to the heating of the dust that is the final source
of most of the observed emission.  However, Veilleux et
al. (1995) have shown in a large sample of infrared galaxies
that the fraction of infrared galaxies with AGN-like optical
spectra increases as a function of increasing luminosity, providing
important circumstantial evidence that AGN-like activity is
associated with ultraluminous infrared galaxies.  The strength
of this evidence might be considered to be weakened, however, by
the finding of Goldader et al. (1995) that the ratio of
Br$\gamma$ to infrared luminosity decreases as a function of
luminosity, suggesting that dust obscuration plays an even
greater role in ultraluminous infrared galaxies than in
luminous infrared galaxies.  This suggests the perverse
possibility that energetically dominant starbursts in
ultraluminous infrared galaxies are so obscured in the optical
that only incidental unobscured AGNs are detected.

Radio observations are unaffected by dust obscuration and can
provide high spatial resolution.  However, the primarily
synchrotron radiation that is detected is only indirectly linked
to the power source of infrared galaxies.  In starbursts, it is
thought that the radio emission arises from electrons
accelerated in supernovae remnants, while in AGN, relativistic
jets are strong radio emitters. Condon et al. (1991; hereafter
CHYT) have employed the high spatial resolution provided by the
Very Large Array (VLA) to show that the radio sizes of a large
fraction of ultraluminous infrared galaxies are comparable to
the minimum size required for a blackbody that reproduces the
25, 60, and 100 \mum flux densities of these sources.  From this
they infer that starbursts rather than AGNs may power these
sources.  At even higher spatial resolution provided by very
long baseline interferometry (VLBI), Lonsdale, Smith \& Lonsdale
(1993) find evidence for radio structure on the scale of
milliarcseconds, some components of which have brightness
temperatures well in excess of what could be expected in
starburst-related emission.  They deduce from this that AGNs are
present in five of eight of the ultraluminous infrared galaxies
in their sample\footnote{Smith, Lonsdale \& Lonsdale (1998)
suggest that radio supernova remnants may be responsible for
some of the structure they detect}.

In this work, the dust emission itself is examined for
spectral clues that can reveal the nature of the power
source in infrared galaxies.  In a sample of 60 galaxies,
Roche et al. (1991; hereafter RASW) have shown that their
8--13 \mum spectra can be classified into three types:
those with prominent polycyclic aromatic hydrocarbon (PAH)
features generally show \ion{H}{2} region-like optical
spectra, those with flat featureless spectra have Seyfert
1 or quasar-like optical spectra, and those rare galaxies
with silicate absorption features usually have Seyfert 2 optical
spectra.  Dudley \& Wynn-Williams (1997) have extended
the work on galaxies with very deep silicate features to
show that the size of the power source can be constrained
to be smaller than a few parsecs, implying that such sources
are powered by deeply embedded AGNs.  Here, the 8--13 \mum
spectra of a well-defined sample of 25 luminous and
ultraluminous infrared galaxies are examined in this
context. 

It should be noted that all of the sources for which new
spectral observations are reported here can be found in one or
more of four {\em IRAS}-based catalogs: the Bright Galaxy Survey
(Soifer et al. 1989), the Extended Bright Galaxy Survey
(Sanders et al. 1995), the Two Jansky Survey (Strauss et
al. 1992), or the Extended 12 \mum Survey (Rush, Makan \&
Spinogio 1993).

Observations and data reduction procedures are described in
Section 2.  Section 3 presents the new spectral observations.  The
classification of the spectra is discussed in
Section 4.1.  The justification for the classification of the ultraluminous galaxies that fall within the later subsample
is discussed in some detail in Section 4.2.  In Section 4.3 a subsample
of galaxies is defined and an apparent trend with luminosity
within it is examined.  The possibility of an
anti-correlation between the equivalent width of the 11.3
\mum PAH feature and the ratio of infrared light to molecular
gas mass is discussed in Section 4.4, and conclusions are
drawn in Section 5.

\section{Observations and Data Reduction\protect\label{ch5:odr}}

The new observations reported here were obtained over the
course of 5 years, from 1991 through 1995, using the 10 and
20 \mum Cooled Grating Spectrograph (CGS3) mounted at the
Cassegrain focus of the United Kingdom Infrared Telescope
(UKIRT) on Mauna Kea in Hawaii.  Observations in 1991 were
obtained by G. Wynn-Williams and J. Goldader, in 1992 by
G. Wynn-Williams and C. Dudley, and in 1993--1995 by
C. Dudley. For parts of each observing run a UKIRT
scientist (J. Davies, T. Geballe, or G. Wright) was present
at the summit and participated in the observations.  CGS3 has
been described by Cohen \& Davies (1995).  All but one of
the spectra reported here were obtained using the low-resolution spectroscopy mode with $\lambda/\Delta\lambda
\sim 60$ in a series of beam-switched, chopped observations
with a chop frequency of $\sim$10 Hz and a beam switch
frequency of $\sim$0.5 Hz.  To fully sample each
spectrum, two interlacing grating positions were observed
with changes between the two positions occurring after
10--16 beam-switched pairs.  Table 1 gives a log of
observations for the galaxies.  In Table 1, column 1 is
the object name, columns 2 and 3 give the aperture centre
in RA and DEC, column 4 gives the position reference,
columns 5 and 6 give the object redshift and reference,
column 7 gives the date of observation, column 8 gives the
circular aperture size used in arcseconds (full width at
10 per cent power), column 9 gives the chop throw in arcseconds,
and column 10 gives the chop direction.

%\begin{figure}
%\plotfiddle{ch5t5.1p1.ps}{\textheight}{180}{100}{100}{300}{710}
%\addcontentsline{lot}{table}{\protect\numberline{5.1}{Observing Log and Assumed Redshifts for Spectroscopic Observations}}
%\end{figure}

%\begin{figure}
%\plotfiddle{ch5t5.1p2.ps}{\textheight}{180}{100}{100}{314}{710}
%\end{figure}

%\begin{figure}
%\plotfiddle{ch5t5.1p3.ps}{\textheight}{180}{100}{100}{314}{710}
%\end{figure}

%\begin{figure}
%\plotfiddle{ch5t5.1p4.ps}{\textheight}{180}{100}{100}{314}{710}
%\end{figure}

%\begin{figure}
%\plotfiddle{ch5t5.1p5.ps}{\textheight}{180}{100}{100}{314}{710}
%\end{figure}

%\addtocounter{table}{1}

Observations were calibrated in wavelength by means of
high order spectral measurements of a Kr lamp in the
manner discussed by Hanner, Brooke \& Tokunaga (1995).
A master calibration spectrum was first selected and
calibrated in wavelength by flattening a 2$\times$
over-sampled Kr lamp spectrum (4 grating positions) with
the spectrum of a soldering iron replacing the Kr lamp in
the beam.  Kr lamp spectra from each night's observations
were then cross correlated against the unflattened master
spectrum to give a wavelength calibration for each night.
Typical shifts from the nominal CGS3 wavelength
calibration were less than a resolution element, and were
never more than 2 resolution elements.

Corrections for instrumental response, atmospheric
transmission, and flux calibration were performed by dividing
source spectra by the spectrum of a standard star with a
known 10 \mum flux density and an assumed blackbody
temperature based on its spectral type, and then multiplying
by the Planck curve of that temperature scaled to match the
standard star in flux density.  Observations of standard
stars were performed with the same aperture, chop throw,
position angle and frequency as the galaxy observations they
calibrate. The standard star observations were made to match
in airmass so that no additional airmass corrections would be
required.  For the standard stars, the number of beam-switched pairs between grating shifts was typically 3--5. The
stars have been chosen, where possible, to have a spectral
type of K0 or earlier to avoid the effects of SiO absorption,
H$^-$ opacity, and circumstellar dust emission that can
effect the spectral shape of late-type stars.  Table 2 gives
the HR number, 10.1 \mum flux density, photometric reference,
and assumed blackbody temperature of each standard star, in
columns 4, 5, 6, and 7, respectively, for each of the
observations given in Table 1.  In column 2 of Table 2,
alternative names are given for galaxies that are found in
multiple catalogs, while column 3 gives the date of each
observation as in Table 1.  It should be noted that an error of $\sim$1000 K in the assumed blackbody temperature of the standard
stars leads to a $\sim$2 per cent error in flux calibration at
either end of the passband 8--13 \mum which is negligible for
the purposes of this work, while for those stars of late
spectral type, deviations from the assumed blackbody can be
$\sim$10 per cent in the SiO band (Cohen et al. 1995).  Systematic
uncertainties in overall flux calibration of 15--20 per cent can
arise due to telescope pointing errors with the stronger
effects for the smaller apertures.  For wavelengths shortward
of 8.2 \micron, strong systematic effects can occur as a
result of a variable water vapor column in the Earth's
atmosphere.

%\clearpage 

All spectra were regridded to a common wavelength scale using a
Gaussian filter of $\sigma$ = 0.064 \mum, which lead to a slight
correlation between neighboring data points.  Those spectra of the
same source obtained on two or more nights were combined,
weighted by integration time.  Data not falling in the ranges
7.7--9.1 and 10.0--13.4 \mum have been excluded due to poor
atmospheric transmission.

Spectra of Wynn-Williams et al.'s (1991) Arp 299B1 were obtained on 1994 February 6 and 7
using CGS3 in high-resolution mode ($\lambda/\Delta\lambda
\sim 250$) using a 3\farcs26 full width at 10 per cent power
circular aperture.  The spectra obtained on each night were
the result of two sets of observations each consisting of two
interlacing grating positions but with different wavelength
ranges to cover the spectral range 10--13.2 \micron.  The chop
throw was 20\arcsec, and the chop and beam switch PA were
45$^\circ$.  About 30 per cent of the first night's observations were
affected by clouds; these data were rejected by inspecting the
change in the dispersion of the total signal.  The Kr lamp was
observed on each night to provide wavelength calibration.  The
strongest lines from each night agreed in wavelength to 3 per cent of
a resolution element so that the Kr spectra of the two nights
were combined, and a difference in wavelength of 30 per cent of a
resolution element was measured between the expected (based on
the nominal calibration of CGS3) and the measured wavelengths
of 5 Kr lines.  An additive correction of 0.02 \mum was
applied to the observations, this being the interpolation of
the 2.117 and 2.191 \mum Kr lines measured at 6th order at the
expected position of the galaxy [\ion{Ne}{2}] line.  Flux
calibration was performed using HR 4067 with a 10.1 \mum flux
density of 103 Jy and an assumed temperature of 4000 K on the
first night, and HR 5340 using a 10.1 \mum flux density of 738
Jy and an assumed temperature of 4400 K on the second night.
No significant difference was found between the two nights'
observations, so they were combined according to their
statistical weights. The difference in airmass between the
source and standard observations was $\sim$0.5 but monitoring
of BS 4518 on the first night at airmasses similar to the
source observations showed that the level of spectroscopic
variation over the observed wavelength range as a function of
airmass was negligible compared to the final signal-to-noise
ratio.  The overall photometric calibration is estimated to be
good to about 25 per cent while the typical signal-to-noise ratio of
an independent data point is about 10.

\section{Results\label{ch5:r}}

The intensity and equivalent width of the 11.3 \mum PAH
feature have been measured directly for all of the sources by
estimating the continuum in the rest wavelength (see Table 1) ranges
10.5--10.9 and 11.7--12.1 \mum and subtracting the linear
interpolation of these two estimates from the spectra to
measure the residual flux in the range 10.9--11.7 \micron.
Uncertainties in the feature intensities represent the
combined uncertainties in the residual flux in this range but
not the uncertainties in continuum subtraction.  For the
equivalent width estimates, uncertainties in the continuum
level have been included.  In most cases the continuum
uncertainties are dominant in the equivalent width estimates,
leading to indeterminate results in the case of limits, but in
a few cases the uncertainty in the residual is the dominant
component, and upper limits to these equivalent widths can be
set.

Since model fitting can lead to substantially cleaner
estimates of feature fluxes (of order the signal-to-noise
ratio in the peak spectral element being fit)
estimates based on the model that is applied to these 
data to test for the presence of a silicate absorption
feature are also given.  The model consists of a
power-law continuum and an additive PAH spectrum
represented by the spectrum of the Orion Bar taken from
Roche (1989).  The fitting of this model was accomplished
using a version of Bevington's (1969) program CURVEFIT
supplied with IDL ver. 4.0 using three free parameters,
the power-law continuum scaling and index ($f_\nu =
\beta\times\nu^{-\alpha}$), and the scaling of the Orion Bar
spectrum.  The data were fit in the rest wavelength ranges
10.0--10.9, 11.0--11.4, and 11.7--12.6 \mum using initial
guesses based on linear fits to the logarithm of data
binned in the ranges 10.0--10.5, 10.5--10.9, and
11.7--12.6 \mum for the power-law scaling and index, and
based on the continuum subtracted estimates of the flux in
the 11.3 \mum feature.  The weighting of the data was
statistical.  As discussed recently by Goldader et
al. (1995), the use of a power-law as a fitting function
tends to throw the bulk of the errors in the fit
parameters into the power-law scaling, so that the formal
errors of the remaining parameters are underestimated.
Further, the model assumptions introduce further
systematic uncertainties.  In particular, by fitting the
data in the range 11.2--12.6 \micron, it is assumed that
the ratio of 11.3 \mum feature emission with respect to the
underlying plateau emission that extends to 12.9 \mum is
the same as that found at the position of peak of the 11.3
\mum emission feature in the Orion Bar.  This ratio,
however, is know to vary spatially within the Orion Bar (Roche
et al. 1989), and also within at least one external galaxy
(Dudley \& Wynn-Williams 1999).  The consequences
of this uncertainty are largest in spectra where the PAH
emission is most powerful relative to the continuum.  Given
that the galaxy redshifts make it impossible to estimate the
continuum beyond 13.1 \mum (rest frame) in $\sim$50 per cent of the
sources, it has not been possible to quantify the importance of
this effect in these data, regardless of obvious
signal-to-noise issues.  Due to these multiple and tangled
uncertainties, both results of direct measurements and model
fits with formal errors are presented, but it must be
cautioned that the model parameter errors are seriously
underestimated.  In particular, model 3 $\sigma$ upper limits
to the 11.3 \mum intensity are roughly the same as measured 1
$\sigma$ upper limits to the 11.3 \mum intensity.  The new
spectra and model fits are presented in Fig. 1.

%\newcount\figcount
%\figcount=0
%\newcount\fpcount
%\newcount\fig
%\fpcount=0
%\def\htr{\ifnum\figcount=0 212 \fi \ifnum\figcount=1 212 \fi \ifnum\figcount=2 440 \fi \ifnum\figcount=3 440 \fi }
%\def\vtr{\ifnum\figcount=0 -270 \fi \ifnum\figcount=1 430 \fi \ifnum\figcount=2 473 \fi \ifnum\figcount=3 1172 \fi}

%\openin1=ch5flist
%\loop \begin{figure} {\loop \message{\the\figcount}  \read1 to \fig  \plotfiddle{\fig}{350pt}{90}{90}{90}{\htr}{\vtr} \advance\figcount by 1  \ifnum\figcount<4 \repeat } \end{figure} \figcount=0 \advance\fpcount by 1 \ifnum\fpcount<8 \repeat 

%\addtocounter{figure}{1}
%\begin{figure}
%\plotfiddle{ch5f1.cap.ps}{\textheight}{180}{100}{100}{300}{710}
%\addcontentsline{lof}{figure}{\protect\numberline{\thefigure}8--13 \mum Spectra of 27 Galaxies}
%\end{figure}

%\clearpage

Table 3 presents the model power-law index in column 2,
the measured (upper) and model (lower) 11.3 \mum feature
intensity in column 3, the 11.3 \mum feature equivalent width in
column 4, the ratio of the flux density of the source in
the rest wavelength range 8.0--8.4 \mum to the model flux
density in the same range where the errors represent the
errors in the estimates of the source flux densities only.
Measurements of fine structure lines are given in columns
6 and 7.  For the measurements of fine structure lines the
continuum was estimated by eye, and for detections of the
12.8 \mum [\ion{Ne}{2}] line it must be cautioned that a
significant amount of the intensity could be due to a
blended PAH emission feature at 12.7 \mum except in the
case of Arp 299B1, where the spectral resolution is
sufficient to distinguish the two features.  Errors for
detections and limits are 1 $\sigma$ and 3 $\sigma$,
respectively, except in the case of the model 11.3 \mum
equivalent width, which is reported in all cases to give a
sense of the model continuum uncertainties.  In the case
of Arp 299B1, the fine structure [\ion{S}{4}] and
[\ion{Ne}{2}] results are derived from the high-resolution
spectrum.  The redshift of the [\ion{Ne}{2}] line (2900
\kms) agrees within 1.4 $\sigma$ (1 $\sigma$ = 170 \kms
based on Gaussian fitting) with the CO redshift measured
by Sargent \& Scoville (1991), while much more precise
agreement between unpublished IRTF CSHELL \brg~ data and the CO
observations is known, so that some
confidence in [\ion{Ne}{2}] identification is warranted. It
should be noted
that [\ion{Ne}{2}] is not detected in the low resolution
8--13 \mum spectrum of Arp 299B1 but that the 3 $\sigma$
upper limit for this feature is 2.1 $\times 10^{-15} $ W
m$^{-2}$, which is consistent with the intensity measured
at high resolution.  Further, the ratio of I(11.3) to
I([\ion{Ne}{2}]) (1.7) is not unusual for galaxies for
which the ratio has previously been measured (RASW).

%\begin{figure}
%\plotfiddle{ch5t5.4p1.ps}{\textheight}{180}{100}{100}{300}{710}
%\addcontentsline{lot}{table}{\protect\numberline{5.3}{Spectral Measurements of Galaxies}}
%\end{figure}

%\begin{figure}
%\plotfiddle{ch5t5.4p2.ps}{\textheight}{180}{100}{100}{314}{710}
%\end{figure}

%\begin{figure}
%\plotfiddle{ch5t5.4p3.ps}{\textheight}{180}{100}{100}{314}{710}
%\end{figure}

\addtocounter{table}{1}

As is apparent in Table 3, while a wide range of power-law
indices are necessary to describe the long-wavelength
continuum, the ratio of 8.2 \mum flux to model 8.2 \mum
flux is not significantly different from unity for a large
fraction of the galaxies even in cases where the power-law
index is large and a minimum near 9.7 \mum is pronounced.
As a result, it is not generally necessary to invoke silicate
absorption to explain the observed spectra.

\section{Discussion\label{ch5:d}}

\subsection{Spectral classification\label{ch5:d1}}

RASW have found that a three-type
classification scheme is sufficient to classify the 8--13
\mum spectra of most galaxies.  Their scheme consists of
flat-featureless, silicate absorption or emission, and PAH-type (their UIR)
spectra.  Further, there is a general correspondence
between the 8--13 \mum spectrum of a galaxy and its optical
spectral classification based on optical emission lines.  The
correspondence is that those galaxies with flat 8--13 \mum
spectra are often broad-line AGNs, those with silicate
absorption are frequently narrow-line AGNs, and those with PAH
features typically have optical spectra with line ratios
similar to Galactic
\ion{H}{2} regions.  

Parenthetically, the identification of the mid-infrared
emission feature analysed here is not entirely secure.  Other
laboratory analogs, including quenched carbonaceous composites
(Sakata \& Wada~1989), hydrogenated amorphous
carbon (Duley~1989), and anthracite (Ellis et
al.~1994), have been proposed in addition to
the PAH model (Puget \& L\'eger~1989) that is
adopted here for ease of expression.  However, the precise
identification of the chemical source of these features is not
particularly important for the present work.  What is
important is that the strength of one feature may be predicted
to some extent from observations of another.  In particular
the strength of the 7.7 \mum feature may be predicted from
observations of the 11.3 \mum feature to within about a factor of 2
(Zavagno, Cox \& Baluteau 1992).

The model fitting performed at the long-wavelength end of the
spectra can be used effectively to test for the presence of
strong silicate absorption.  Two cases might occur.  First, a
PAH spectrum might undergo external extinction.  In this case,
two effects will contribute to the model under predicting the
observed flux in the shorter wavelength half of a spectrum.  The
first effect is that the 11.3 \mum feature will undergo more
severe extinction than the region near 8 \mum due to the
spectral shape the silicate component of the interstellar
extinction curve.  This has the effect of reducing the predicted
PAH strength.  The second effect is that the slope of the model
continuum will have a larger spectral index than would otherwise
be the case if no silicate absorption were present.  The model
would therefore under-predict the continuum underlying the 7.7
and 8.6 \mum feature complex.  Both of these effects will cause
the model to under-predict the shorter wavelength observations.
Second, a PAH spectrum may undergo no particular extinction,
however, a second silicate-absorbed continuum component may also
contribute to the spectrum.  In this case, the model should
predict the PAH component well.  However, if the silicate
absorption is strong, the model spectral index will be increased
due to the additional component, and the continuum in the short
wavelength half will be underestimated so that the model will
also under-predict the observed flux.  

The RASW classification scheme is applied to the present data
in the following way: the ratio of the observed to model flux
density between 8.0 and 8.4 is used as a guide to classification
such that if this ratio is 3 $\sigma$ above 2.5, then the
source is classified as silicate.  This cut may somewhat
underestimate the number of galaxies in which silicate
absorption is present compared to the approach taken in
Dudley \& Wynn-Williams (1999), where the cut is taken at
2.0. For the purposes of definite classification, under the
working hypothesis that all infrared galaxies are PAH-type
unless convincingly shown to be otherwise, this more
conservative value is appropriate.  In cases where the 11.3
\mum feature is not formally detected but only allowed (dot
dashed curves in Fig. 1), and where a minimum is present
near 9.7 \micron, a question mark is appended to the PAH
designation; however, whether or not there is a formal
detection of the 11.3 \mum feature in sources with obviously
flat spectra, the classification is given as flat.  A mixed
classification of PAH and silicate is given where PAH
emission can explain $\sim$30 per cent of the observed emission
between 8.0 and 8.4 \micron.  Under this condition, the
classification of Arp 220 by RASW as both PAH and silicate
would be revised to just silicate.  Arp 220 is discussed
in detail in Section 4.2.5.

%\begin{figure}
%\plotfiddle{ch5t5.6p1.ps}{\textheight}{180}{100}{100}{300}{710}
%\addcontentsline{lot}{table}{\protect\numberline{5.4}{Spectral Classification and Other Observables}}
%\end{figure}

%\begin{figure}
%\plotfiddle{ch5t5.6p2.ps}{\textheight}{180}{100}{100}{314}{710}
%\end{figure}

%\begin{figure}
%\plotfiddle{ch5t5.6p3.ps}{\textheight}{180}{100}{100}{314}{710}
%\end{figure}

\addtocounter{table}{1}

%\clearpage

Table 4 gives, for each source in column 1, the spectral
classification based on the present data and on optical
observations, along with 12--100 \mum photometry, the infrared
luminosity, and the star formation efficiency.  In some cases
the optical spectral classification has been inferred from the
data of Veilleux et al. (1995), if some diagnostics were
missing. It is clear from Table 4 that the general
correspondence between optical spectroscopy and 8--13 \mum
spectroscopy is quite similar to that found by RASW.  Neglecting
LINER-type galaxies, four exceptions where AGNs are identified
optically but which are found to have PAH-type spectra in this
work are Mkn 1034, IRAS 04154+1755, IRAS 05189$-$2524, and Zw
475.056.

Mkn 1034 is a known Seyfert 1 galaxy; however its 8--13 \mum
spectrum can be fit with a PAH plus continuum model.  This may
be consistent with the finding of Sopp \& Alexander (1992) that
only half of the 6-cm flux density measured in a 14\farcs5 beam
is also contained within a 1\farcs5 beam, suggesting that
spatially extended star formation plays a role in this source.
A similar explanation may be invoked for the presence of PAH
emission in the spectra of the Seyfert 2 galaxies IRAS
04154+1755 and Zw 475.056.  Crawford et al. (1996) find IRAS
04154+1755 to be resolved at 6 cm using both 4-arcsec and
0.5-arcsec beams.  In the map of Zw 475.056 presented by CHYT,
low-level radio emission extends over 1.5-arcsec.  IRAS
05189$-$2524 is discussed in detail in the next section; however
it should be noted here that the presence of exceptions, in this
sample, to the the correspondence between optical and 8--13 \mum
spectral classification found by RASW should not be taken as
evidence that the correspondence is merely coincidental.  The
sources presented by RASW were sometimes selected on the basis
of optical spectroscopy to examine the question of what kind of
spectrum they would present at 8--13 \micron.  Hence, many of
their sources were archetypical examples of their optical class.
In the present sample, selection is based solely on infrared
properties, so that mixed cases might well be expected.

It is probably not coincidental that these four sources are
among the sources in Table 3. with the smaller values of
$\alpha$ suggesting that the AGN in these sources may contribute
somewhat to the 8--13 \mum continuum; strong 12 and 25 \mum
emission is known to select for AGNs that have unambiguous
optical spectra (de Grijp et al. 1985; Spinoglio \& Malkan, see also
the AGN spectral energy distributions presented by Sanders et
al. 1989).

\subsection{Comments on individual ultraluminous infrared galaxies}

%\begin{figure}
%%\plotfiddle{ch5f2.eps}{265pt}{0}{60}{60}{-231}{0}
%\caption[IRAS 05189-2524: Starburst or AGN?]{The spectral energy
%distribution of IRAS 05189-2524 is compared with those of
%another ultraluminous infrared galaxy: Mkn 231 and the
%infrared loud quasar IRAS 13349+2438.    In
%the figure, 0.9--5 \mum photometry (open diamonds) are from
%McAlary, McLaren, \& Crabtree (1979) for Mkn 321, from Dudley
%(1998) for IRAS 05189-2524 with the exception of the 0.9 \mum
%point which is from Surace (1998), and from Beichman et
%al. (1986) for IRAS 13349+2438.  8--13 \protect\mum spectroscopy
%(filled circles) is from Roche, Aitken \& Witmore (1983) for Mkn
%231 and from this work for the remaining two sources.  IRAS data
%are given by open squares.  The data for IRAS 13349+2438 have
%been presented in the quasar's rest frame and the spectral
%energy distributions of IRAS 05189-2524 and IRAS 13349+2438
%have been divided by factors of 2 and 20 respectively.}
%\end{figure}

The observed properties of the ultraluminous infrared galaxies
presented here show a surprising range.  A closer look at these
properties will be discussed in this section.  

\subsubsection{A phenomenological description of IRAS 05189$-$2524}

One of the greatest surprises in these data is the case of
IRAS 05189$-$2524, which might have been expected to show a flat
or silicate absorbed 8--13 micron spectrum based on its
optical AGN-like spectrum (Sanders et al. 1988).  What is
observed however is a spectrum that is well fit by a PAH plus
power-law continuum model (see Fig. 1).  Additionally, the
strength of the 11.3 \mum emission relative to the 8--1000
\mum emission is quite similar to known starburst galaxies so that
it seems possible that star formation might be the primary
power source of the galaxy.  Here, a tentative decomposition
of its spectral energy distribution is proposed.  In Fig.
2, a comparison is made between IRAS 05189$-$2524 and two
known AGNs to suggest that while the 1--5 \mum emission in IRAS
05189$-$2524 may be due to hot dust heated by an AGN, the 8--100
\mum emission may be better explained as arising from star
formation traced by the observed PAH emission.  In Fig. 2
the power-law extrapolations of the dust continuum emission
(solid lines) found in the 8--13 \mum spectra are projected to 5
and 25 \micron.  For the two sources where the 8--13 micron
emission is thought to be due to AGN-heated dust (Mkn 231 and
IRAS 13349+2438) (Roche et al. 1983; Beichman et al. 1986),
these extrapolations are reasonable predictors of the observed
flux densities at both 5 and 25 \micron.  For IRAS 05189$-$2524
the extrapolation of the continuum to 25 \mum is in reasonable
agreement with the observed flux density; however the
extrapolation to 5 \mum fails to predict the 4--5 \mum
photometry of IRAS 05189$-$2524 (see double arrow), suggesting
that an unrelated power source is responsible for this
emission.  Given the apparent similarity in the shape of 1--5
\mum emission in both IRAS 05189$-$2524 and IRAS 13349+2438, it
is possible that the spectral energy distribution of IRAS
05189$-$2524 is the result of a starburst producing the majority
of the 8--100 \mum emission and an AGN with a spectral energy
distribution similar to that of IRAS 13349+2438 responsible
for the bulk of the shorter wavelength emission but
contributing little to the 8--100 \mum emission.  Such a
decomposition could reconcile the apparently contradictory
observations of both an AGN-like optical spectrum and the
present PAH type 8--13 \mum spectrum.  A decomposition of this
kind could also be consistent with the suggestion of Goldader
et al. (1995) that the 2 \mum continuum is dominated by AGN
[heated dust] emission based on the small equivalent width of the observed
\brg~ emission.  That hot dust typically plays a lesser
role in starburst galaxies at 2 \mum may also be deduced
from the frequent detection of stellar continuum with probably
little veiling of the CO absorption features, while in IRAS
05189$-$2524 CO absorption is not detected (Goldader et
al. 1995).  Another piece of evidence that supports the view that
the AGN in IRAS 05189$-$2524 may not be energetically dominant
is the failure to detect a radio core of extreme brightness
temperature in VLBI observations (Lonsdale, et al.
1993).

The VLA radio map of CHYT shows the radio continuum in this
source to be extended on a scale of 160 pc, which is
comparable to the minimum blackbody diameter of 280 pc based on
the observed 60 and 100 \mum flux densities and assuming optically thick
dust emission.  It is therefore plausible that the far-infrared
emission arises from optically thick warm dense clouds embedded
in the starburst with a low volume filling factor but a column
covering factor of order unity. The mid-infrared continuum and
PAH emission present in this spectrum might arise from the
surfaces of these clouds and may be optically thin in the sense
that the emission arises from cloud surfaces facing our line of
sight, and from a layer that is only 1--2 $A_V$ thick on any
given line of sight into the starburst.  

Given the possibility
of a decomposition of the 1--100 \mum spectral energy
distribution into a dominant starburst component and an AGN
component important only at shorter wavelengths, the
plausibility of the infrared emission being spatially associated
with the radio continuum, and most importantly, the presence of
11.3 \mum PAH emission in the 8--13 \mum spectrum taken together with the
adequate fit of the model given in Fig. 1, the 8--1000 \mum
emission from IRAS 05189$-$2524 is attributed to a starburst for
the purposes of this work.

\subsubsection{Digression on newly available {\em ISO} data}

The model applied to the data presented in Fig. 1 for the
purpose of spectral classification is not required in cases
where the strength of the 7.7 \mum PAH feature has been
observed as is now the case for four of the ultraluminous
infrared galaxies in this combined sample as a result of {\em
Infrared Space Observatory (ISO)} observations (Genzel et
al. 1998).  With the additional information it is possible to
compare observations with a variety of scenarios to better
understand the physical situation with greater confidence than
possible when only ground-based data are available.  In Fig. 3,
three possibilities are considered: (1) For the top spectrum
in each panel, a PAH plus power-law continuum model such as
that employed in Fig. 1 is fit not just in the spectral region
surrounding the 11.3 \mum feature, but to the entire 6--13 \mum
range.  (2) For the middle spectrum in each panel, a PAH plus
power-law continuum model is subjected to extinction by cold
dust that includes the silicate absorption feature, as adopted
by Dudley \& Wynn-Williams (1997), and again fit to the entire
6--13 \mum range.  (3) For the bottom spectrum in each panel, a
PAH plus power-law continuum model is additively combined with
a silicate-absorbed continuum source, and again fit over the entire
6--13 \mum range.  The model fits presented in Fig. 3 have been
carried out using CURVEFIT with equal weighting.  In the first
model, the index of the power-law was specified either from
Table 3 (IRAS 17208$-$0014 and Mkn 273) or as 12.8 and 4.9 for
Arp 220 and Mkn 231, respectively.  It should be noted that
three of the sources exhibit a large dynamic range, and in the
following discussion some attention will be drawn to the poor
fit at low flux density levels, which is also an expected systematic
effect of the fitting procedure, since differences between
small numbers are small.  It may also be noted that the
mismatches are apparent for a large number of neighboring data
points.

\begin{figure}
%\plotfiddle{ulirgplot1.eps}{400pt}{0}{61}{61}{-225}{0}
\end{figure}

%\begin{figure}
%%\plotfiddle{ulirgplot2.eps}{400pt}{0}{61}{61}{-225}{0}
%\caption[Dust emission in Ultraluminous Infrared Galaxies]{The
%combined ground based and {\em ISO} spectra of 4 ultraluminous
%galaxies are presented.  For each galaxy, the top spectrum
%gives the observed spectrum, while the lower two are shifted
%by multiplicative factors.  The ground based data (filled
%circles) are from this work for IRAS 17208-0014 and Mkn 273,
%from Smith et al. (1989) for Arp 220, and from Roche et
%al. (1983) for Mkn 231.  The {\em ISO} data (open circles) are
%from Genzel et al. (1998) and have been corrected by a factor
%of 1.4.  Three dust emission models are shown for each source
%with the total emission given by the solid lines and continuum
%emission given by the broken lines.}

%\end{figure}

%\clearpage

\subsubsection{IRAS 17208$-$0014}

In IRAS 17208$-$0014, all models give fairly adequate fits
to the data, and deciding between them depends mainly on
other observational evidence.  The strongest evidence that
supports a preference for the fit shown for the top spectrum
is that this source is extended on a scale $\sim$1.7 kpc in
the radio (Condon et al. 1996), although higher spatial
resolution observations reveal the presence of a sub-kpc
radio continuum source associated with the OH maser emission
(Martin et al. 1989).  The failure of Martin et al. (1989)
to detect more extended emission may be due to the
restricted set of base lines they employed.  When this
evidence is taken along with the optical spectral
classification of Veilleux et al. (1995) in which the line
ratios in this source are similar to those in \ion{H}{2}
regions, there is no strong reason to suspect that anything
other than star formation is responsible for the observed
mid-infrared spectrum.  Nor is there any reason to suspect a
high degree of extinction; IRAS 17208$-$0014 is among the
sources presented by Genzel et al. (1998) for which the
continuum underlying the 12.8-\mum [\ion{Ne}{2}] feature in
their SWS data can be comfotably plotted on the scale of the
ISOPHOT-S data in their fig. 2, suggesting a lack of strong
silicate absorption (see also Fig. 1).  The fit to the upper
spectrum might be improved somewhat by increasing the
strength of the 7.7 and 6.2 \mum features by a factor of 2,
a procedure which is probably allowed for the 7.7 \mum
feature at least since the Orion Bar spectrum used as a
template here is known to have a somewhat weaker than usual
7.7 \mum feature relative to the 11.3 \mum feature (Zavagno
et al. 1992).

\subsubsection{Mkn 273}

The fit to the top spectrum of Fig. 3 for Mkn 273 gives, as
might have been anticipated from Fig. 1, a poor representation
of the data.  The fitted strength of the 11.3 \mum feature is
clearly too large, which is roughly the converse of the result
in Fig. 1.  Thus, it may be concluded that a simple PAH plus
power-law continuum model does not describe these data.  The fit
shown for the middle spectrum is better, in that it requires
less PAH emission over all and the strength of the 11.3 \mum
feature is not over-predicted; however the continuum around the
9.7 \mum silicate minimum is not well reproduced.  This
suggests that an unaccounted-for emission component is required
to produce this emission.  The fit to the lower spectrum
overcomes this difficulty by superimposing an unobscured PAH plus power-law
continuum such as that which fit IRAS 17208$-$0014 (top
spectrum) upon a continuum source that suffers strong silicate
absorption.  Since Mkn 273 shows clear signs of nuclear
activity (Armus, Heckman \& Miley 1989; Veilleux et al. 1995),
a silicate-absorbed continuum source might be produced by an
edge-on view of a disk model such as those proposed by Pier \&
Krolik (1992) or Efstathiou \& Rowan-Robinson (1995), which are
natural elaborations of the AGN unification model proposed by
Antonucci \& Miller (1985).  The continuum level evident in the
SWS observations of 12.8-\mum [\ion{Ne}{2}] in Genzel et
al. (1998) is consistent with the new CGS3 data presented here;
and therefore consistent with silicate absorption playing a
role in this source.  If the fit to the bottom spectrum is accepted,
then the AGN component could well be responsible for the
greater fraction of the 60 and 100 \mum emission.

\subsubsection{Arp 220}

Prior to the new {\em ISO} observations the analysis of the
Arp 220 spectrum would have proceeded very much along the
lines that have just been given for Mkn 273 (Smith et
al. 1989), with the proviso that Arp 220 has less obvious
signs of activity in its nucleus so that disk-like models
for the silicate-absorbed continuum emission may be less
helpful in advancing our understanding, and deeply embedded
models perhaps more descriptive (Dudley \& Wynn-Williams
1997).  Applying the model used in Fig. 1 to the data of
Smith et al. (1989) severely under-predicts the observed 8.2
\mum flux density.  However, as can be seen in Fig. 3 the
new {\em ISO} data can apparently be fit by a simple PAH
plus power-law continuum model due to the apparent
discrepancy between the observations of the 11.3 \mum
feature reported by Smith et al. (1989) and the {\em ISO}
data of Genzel et al. (1998).  It should be noted that there
are no continuum points longward of the 11.3 \mum feature in
the new {\em ISO} data, so that determining its strength
from those data is difficult.  It seems safer at this point
to accept the ground-based data as a guide in this spectral
region and thus hold this model in abeyance, pending further
observational evidence.  The fit to the middle spectrum
is essentially the same as that given in fig. 1 of Smith et
al. (1989) and suffers from the same difficulty noted by
Smith et al. (1989), namely that the minimum near 9.7 \mum
is poorly fit.  The fit to the lower plot is similar to that
shown in Smith et al.'s (1989) fig. 2 or as adopted by Dudley \&
Wynn-Williams (1997).  The continuum in the 12.8-\mum
[\ion{Ne}{2}] SWS data of Genzel et al. (1998) is consistent
with the Smith et al. (1989) data but is not apparently
consistent with the ISOCAM-CVF spectrum presented by Elbaz
et al. (1998).

For Arp 220, therefore, the new {\em ISO} data presented by
Genzel et al. (1998) seems to allow the possibility that
unextinguished PAH emission plus a simple power-law
continuum describes this spectrum as a result of the
apparent strength of the 11.3 \mum feature in the new data.
It has been suggested that differing spectral resolution
(roughly a factor of 2) between the Smith et al. (1989) and
Genzel et al. (1998) data may account for this difference,
however this could only work and if the 11.3 \mum feature
were unresolved in both spectra, and that does not appear to
be the case.  Gauging the over all PAH emission from the 6.2
\mum feature which displays continuum on either side, the
intensity of which is found to be well correlated with the
that of the 7.7 \mum feature in Galactic sources (Cohen
et al. 1989), it is clear the PAH emission relative to the
over-all 8--1000 \mum emission is exceedingly weak, really
no stronger than what might be allowed in Mkn 231 (see
next), if the apparent 6.2 \mum feature in the {\em ISO}
data for Mkn 231 is real.  In contrast, again using the 6.2
\mum PAH feature as a guide, the strength of the PAH
emission relative to the 8--1000 \mum emission in Mkn 273 is
twice as strong, and in IRAS 17208$-$0014, five times as
strong. The Arp 220 spectrum might possibly be produced by a
starburst that reprocesses more that 90 per cent of its
luminosity not once but many times so that the observed PAH
emission is produced only in a thin outer shell or in a
small number of narrow chinks. Such a model would be even
more extreme than what is argued above as plausible for IRAS
05189$-$2524 in a starburst context.  However, given the
disagreement between the two spectra of this source near
11.3 \mum and in consideration of the arguments of Dudley \&
Wynn-Williams (1997), a model that has both unobscured PAH
emission and a silicate-absorbed continuum due to a deeply
embedded AGN continues to seem more natural. This
interpretation also seems to account simultaneously for both
the 158-\mum [CII] (Luhman et al. 1998) and 2-10 keV X-ray
(Iwasawa 1999) deficits in this source if the ``AGNs''
(Soifer et al. 1999) are sufficiently obscured.  It should
be noted that in this preferred model, only a fraction of
the flux density at 7.7 \mum can be attributed to the 7.7
\mum PAH feature, in accordance with either the 12.8 of 14.3
\mum continuum seen in the SWS data presented by Genzel et
al. (1998).  As suggested by Dudley \& Wynn-Williams (1997),
spatially resolved spectroscopic observations may help to
clarify this issue.

\subsubsection{Mkn 231}

For Mkn 231 any PAH contribution to the spectrum must be
minimal compared to a continuum contribution (Roche et
al. 1983), as can also be seen from the fit to the upper
spectrum in Fig. 3.  This source had been fit adequately
with a model consisting of a continuum source suffering
silicate absorption by Roche et al. (1983), and the present
models are {\it not} an improvement.  It would be of
interest to know if disk radiative transfer models such as
those proposed by Pier and Krolik (1992) would fair as well
as the cold absorber model of Roche et al. (1983) in the
detailed fit to the silicate absorption feature given their
likely success in fitting the overall infrared spectral
energy distribution of Mkn 231.  The 11.3 \mum emission
feature is not detected in this source and it is not clear
that the single point that is spectrally coincident with the
6.2 \mum feature in the {\em ISO} data should be considered
as a real indication of the presence of PAH emission.  Some
PAH emission is not unexpected in this source given the
presence of star-forming knots near the nucleus of this
source that might account for $\sim$10 per cent of its
infrared luminosity (Surace et al. 1998).  The estimate of
the 7.7 \mum PAH strength given by Genzel et al. (1998)
relies on a continuum point at 10.9 \mum which lies in the
silicate absorption feature.  Of the sources measured in
this way by Genzel et al. (1998) Mkn 231 perhaps provides
the clearest example of how this method is prone to
overestimate the PAH strength when silicate absorption is
present.  The 12.8 or 14.3 \mum continuum in their fig. 1
provide a rough guide as to when there may be a problem in
this respect.  For Mkn 231, the 12.8 \mum continuum of
Genzel et al. (1998) is high in comparison with the data of
Roche et al. (1983) but it is possible that it would agree
with the extrapolation of the long-wavelength ISOPHOT-S data
which also apparently fall significantly above the Roche et
al. (1983) data as seen in Fig. 3.  It is not clear if these
differences should be attributed to aperture effects, or
other systematics, or if they reflect intrinsic source
variability over 1.4 decade; Roche et al. (1983) found good
agreement with the narrowband photometric observations of
Reike (1976) made over 0.4 decade earlier.

\subsection{A 60 \protect\micron~ flux-limited subsample\label{ch5:d2}}

%\subsubsection{sample definition}

Among galaxies with \lir $\geq 1.6\times10^{11}$ \lsun, a nearly
complete (all but three sources have 8--13 \mum spectra)
flux-limited (at 60 \micron) sample of 25 sources may now be
defined on the combined Bright Galaxy Survey (Soifer et
al. 1989) and the Extended Bright Galaxy Survey (Sanders et
al. 1995) (combined BGS) for $F_\nu$(60) $\geq$ 11.25 Jy
(roughly twice the flux limit of  combined BGS).  The only other
restriction on this subsample is that the sources be observable
using UKIRT, i.e. $-25^\circ$ $\leq \delta \leq $ 60$^\circ$.
Table 5 gives the ultraluminous and luminous samples with the
source name in column 1, the 8--13 spectral classification in
column 2, the reference for the 8--13 \mum spectrum in column 3,
the $\log$ of the 8--1000 \mum luminosity in column 4, and the
$\log$ of the ratio of the 12 \mum flux density to the 60 \mum
flux density in column 5. The ultraluminous sample has 5 members
and the luminous comparison sample has 20 members.  The
ultraluminous sample is {\em unlikely} to span all the possible
characteristics of ultraluminous infrared galaxies and therefore
is not fully representative.  There are two observations that
confirm this: (1) if Arp 220 were scaled to the 60 \mum 11.25 Jy
flux limit used to define this sample, it would not have been
observable using CGS3 or other ground-based spectrographs
available when these data were collected, and (2) the ratio
given in column 5 of Table 5 for the two brightest ultraluminous
galaxies (Arp 220 and Mkn 231) bracket the same ratio for all
the galaxies in the sample with $1.6\times10^{11} \leq$ \lir
$\leq 10^{12}$ \lsun~ except NGC 1068 and NGC 7469.  An analysis
over the entire Bright Galaxy Survey shows that the dispersion
of this ratio is indeed larger for ultraluminous infrared
galaxies than for luminous infrared galaxies by a factor of 2
(Soifer \& Neugebauer 1991).  Thus, conclusions based on the
present flux-limited sample of ultraluminous galaxies should be
viewed with caution not simply because of the small number of
galaxies in this sample (counting statistics on a bimodal parent population), but because the
characteristics of the parent population are not well
represented (under-sampling).

It is possible that the last of these two difficulties
could be somewhat ameliorated by the following
considerations.  Ultraluminous galaxies that fall outside
the range of 12 on 60 \mum flux density ratios defined by
Arp 220 and Mkn 231 may be assigned to the AGN-dominated
bin in the context of exploring the AGN vs. starburst
hypotheses.  For ultraluminous galaxies with 12/60 \mum
colors redder than Arp 220, PAH emission cannot make a
contribution to the 12 \mum flux density that is
sufficiently large to explain the 60 \mum flux density. 
It is therefore likely that such sources will have deep
silicate absorption features, and by the arguments
presented by Dudley \& Wynn-Williams (1997), the size of
their power source should be too small to be due to a
starburst.  For galaxies with 12/60 \mum colors bluer than
Mkn 231, 60 \mum selected samples will overlap 25 or 12
\mum selected samples that are known to select for AGNs
(de Grijp et al. 1985; Spinoglio \& Malkan 1989).  However,
reliance on these notions would be ill advised given the
internal evidence in the present data showing that PAH
emission can be important in ultraluminous galaxies with
12 to 60 \mum flux density ratios similar to the two
extremes, namely, IRAS 17208$-$0014 and IRAS 05189$-$2524.

%\subsubsection{trend with luminosity}

Inspection of Table 5 shows that 3 out of 5 of the ultraluminous
galaxies are powered predominantly by AGNs although if judgment
is reserved for Arp 220, this fraction would be 2 out of 5. The
trend is thus consistent with about half of ultraluminous
galaxies being powered by AGNs.  On the other hand, the fraction
of the more numerous luminous galaxies in this sample that are
powered predominantly by AGNs is at most 0.25 and could be as
small as 0.05 or smaller if the relative contributions of the
AGN and starburst in NGC 1068 are approximately equal (Telesco
et al. 1984).  This trend supports the evidence from optical
spectroscopy that AGNs are present in greater numbers in
ultraluminous infrared galaxies (Veilleux et al. 1995).  It goes
further than optical or radio studies because it is possible to
demonstrate with greater certainty which type of power source
(starburst or AGN) dominates the infrared luminosity.  Further,
it is apparent that for luminosities less than 10$^{12}$ \lsun,
star formation is far and away the dominant power source, a
result that confirms work at many wavelengths.  This analysis
lends some support to the proposal of Sanders et al. (1988) that
there is an evolutionary connection between ultraluminous
infrared galaxies and quasars, while at the same time
demonstrating that massive star formation can indeed be the
dominant power source of some ultraluminous infrared galaxies.
However, it must be conceded that the evidence presented
constitutes only a trend. The precise proportion of
AGN-dominated sources among 60 \mum selected ultraluminous
infrared galaxies cannot be said to have been established here,
and thus no statistically meaningful difference between luminous
and ultraluminous galaxies has truly been shown.  It is
of interest that the trend does continue with the next step down
in 60 \mum flux density.  Adding IRAS 12112+0305 and IRAS
08572+3915 to the present sample of five ultraluminous galaxies adds
one apparently starburst-dominated (based on the apparently
normal ratio of the 6.2 \mum PAH feature emission to overall infrared
emission in the IRAS 12112+0305 spectrum of Genzel et
al. 1998\footnote{A crude attempt to correct for the
contribution of the silicate absorption edge to the 7.7 \mum PAH
feature by fitting a power law to continuum points at 5.9 and
14.3 \mum suggests a deficit in the strength of the 7.7 \mum feature for
IRAS 12112+0305.})
and one AGN-dominated object (Dudley \& Wynn-Williams 1997).

\subsection{Anti-correlation between 11.3 \mum eqivalent width and
$L_{\rm IR}$/$M({\rm H}_2)$\label{ch5:d3}}

%\begin{figure}
%%\plotfiddle{ch5f3.eps}{350pt}{0}{120}{120}{-231}{0}
%\caption[Star Formation Efficiency Verses 11.3 \mum EW]{A possible anti-correlation between star
%formation efficiency and the equivalent width of the 11.3
%\mum PAH feature is shown.}
%\end{figure}

In Fig. 4  a possible anti-correlation between
the equivalent width (EW) of the 11.3 \mum feature and the global ratio of $L_{\rm IR}$/$M({\rm H}_2)$ is presented for a number of galaxies.  The
EW data are based either on the
model fits given in Table 3 or the previously published
values of RASW reproduced in Table 6. The ratios of
infrared luminosity to H$_2$ mass are taken from the
literature (Tables 4 and 6).  The solid
line in Fig. 4 indicates inverse proportionality.

The ratio ${L_{\rm IR} \over {\rm M(H}_2)}$ is often taken to be
a measure of star formation efficiency.  When it is large, then
a large fraction of the available gas is involved in star
formation; when it is small, conditions may be such that the gas
exists in a more quiescent state.  However, for a very large
ratio of ${L_{\rm IR} \over {\rm M(H}_2)}$, Sanders et
al. (1991) argued that even a very high star formation
efficiency cannot account for the infrared luminosity, on the
grounds that star formation would too quickly consume the
available gas mass, so that AGN-like activity might be the more
likely primary power source of the infrared luminosity.  Both
${L_{\rm IR} \over {\rm M(H}_2)}$ and EW(11.3) are independent
of distance in so far as EW(11.3) is independent of aperture
size.

A tendency for EW(11.3) to be anti-correlated with the star
formation efficiency might arise if two conditions were met.
First, if the intensity of the 11.3 \mum feature is a roughly
constant fraction of the overall infrared emission (proportional
to $L_{\rm IR}$) then this will not contribute to the
anti-correlation.  Second, if the strength of the continuum at
11.3 \mum increases with decreasing availability of dust
(inversely proportional to M(H$_2$)) to reprocess the light
emitted by massive stars so that a larger fraction of the dust
grain population is heated to the warm temperatures required for
the dust grains to emit at 11.3 \mum then this could be the
source of anti-correlation.  Such an explanation requires
further elaboration since the process by which the 11.3 \mum
continuum is produced is likely, in general, to be a local
phenomenon rather than a global one.

An elaboration that may be in keeping with recent {\em HST}
observations (Watson et al. 1996; Surace et al. 1998;
Satyapal et al. 1997) is that as the star formation
efficiency increases, the proportion of massive stars formed
in dense clusters may increase.  If this is the case, then
the time-scale for clearing the interstellar medium from
around these clusters to a distance where dust grains would
no longer be heated sufficiently to contribute to the 11.3
\mum continuum may be longer that the time-scale for
clearing the interstellar medium around the groups of
massive stars that might form if the star formation
efficiency is lower.  Two effects may contribute to this.
First, the distance at which a grain could be heated to a
temperature high enough to contribute to the 11.3 \mum
continuum would increase as the square root of the number of
massive stars in a cluster so that if the velocity of gas
and dust being cleared from clusters of massive stars is not
too different from the velocity of gas and dust cleared from
groups of massive stars, the dust remains heated for a
longer period of time.  Second, dense clusters might be
temporarily gravitationally bound and behave ballistically
with respect to the dissipative gas. They may sometimes move
into a fresh reservoir of gas, heating the associated dust
to higher temperatures.

In this view, the additional contribution to the 11.3 \mum
continuum emission at higher star formation efficiency would
be provided by possibly equilibrium temperature grains
heated by a combination of direct and Ly$\alpha$ heating
rather than small grains heated by a single photon.  To
zeroth order, this is a sensible proposition, since one
expects (at least within photo-dissociation regions) that
the PAH emission and the spike-heated grain emission to have
roughly fixed relative contributions at 11.3 \mum due to the
rough similarity of the emission time scales for the two
emission processes, so that the apparent anti-correlation
would not necessarily arise.

A number of observations will be required to examine this
situation.  Most important are higher signal-to-noise
observations that will allow real measurements of EW(11.3)
for a number of the sources included in Fig. 4.  These are
required to decide if the anti-correlation is has any merit.
Another avenue of investigation would be to make higher
spatial resolution observations of nearby starbursts where
clusters of massive stars are observed to exist to establish
whether or not EW(11.3) does decrease in the vicinity of
some clusters (see Dudley \& Wynn-Williams 1999 for a
preliminary attempt).  Finally, high spatial resolution 2
\mum adaptive optics imaging of a number of starbursts with
a range of star formation efficiency can help to determine
if the fraction of massive stars that exist in dense
clusters varies with the star formation efficency parameter
as conjectured here.

\section{Conclusions\label{ch5:c}}

New 8--13 \mum spectroscopy of 27 infrared galaxies has been
presented.  The spectra have been fit with a two-component model
consisting of a power-law continuum underlying a PAH emission
spectrum.  These fits are based on the spectral data between 10 and
13 \mum and are used to predict the flux
emitted between 8.0 and 8.4 \micron.  It is
found that for the majority of the observed galaxies, the
model predicts the observed 8.0--8.4 \mum emission to
within an acceptable factor of 2, suggesting that the
majority of the sources are powered by starbursts.

One source, Mkn 273, has been found to show very clear evidence for
silicate absorption.

A flux-limited 60 \mum selected subsample that combines the
present spectroscopic results and those of RASW had been
defined, consisting of 25 galaxies.  In this subsample the fraction of ultraluminous
infrared galaxies powered by AGN is $\sim$50 per cent in
contrast to 5--25\% among the luminous infrared galaxies.  
Owing to the small sample size for the ultraluminous infrared
galaxies, this difference in AGN fraction can be considered only
a trend.

A possible anti-correlation is identified between the
equivalent width of the 11.3 \mum PAH feature and the star
formation efficiency in galaxies where both have been
measured.  It is suggested that variation in the continuum
dust temperature as a function of star formation efficiency
could produce this apparent effect.

\noindent{ACKNOWLEDGMENTS}

The author would like to thank the UKIRT staff: Joel Aikock,
Tim Caroll, Dolores Walthers, and Thor Wold, for assistance at
the telescope, and John Davies, Tom Geballe, and Gillian Wright
for fruitful discussions and calibration data in advance of
publication.  UKIRT is operated by the Joint Astronomy Centre
on behalf of the U.~K.~Particle Physics and Astronomy Research
Council.  Brian Rush, Reinhart Genzel, Vassilis Chamanadaris,
Joe Hora, and D.-C. Kim all kindly provided data in advance of
publication. A series of exchanges with Dieter Lutz has help to
clarify certain issues, as has a discussion with Gary
Neugebauer, Keith Matthews and Eichi Egami.  Eric Becklin has
provided thoughtful comments at several stages of this work.
Many members of the Institute for Astronomy have benefited this
work through discussion and sharing of work in progress.  A few
are Gareth Wynn-Williams (dissertation committee chair), Dave Sanders, Bob Joseph, Klaus
Hoddap, Josh Barnes, Alan Tokunaga, Jeff Goldader, Jason
Surace, and Masatoshi Imanishi; thanks are due to these and
others.  Louise Good proofread the manuscript prior to
submission.  This work has been partially supported by NSF
grants ASTR-8919563, NASA grant NAGW-3938, and the Frank Orrall
fund.  The NASA Extragalactic Database (NED) has been a
constant aid.  It is run by the Jet Propulsion Laboratory,
California Institute of Technology, under contract with NASA.
This work has also made use of NASA's Astrophysics Data System
Abstract Service.

\end{document}